\documentclass[aps,prb,preprint,groupedaddress,eqsecnum]{revtex4-2}

\usepackage{bm}
\usepackage{amsmath}
\usepackage{amssymb}
\usepackage{graphicx}
\begin{document}
\def\opf{\Omega}
\def\opfe{{\overline{\opf}}}
\def\khat{\hat{k}}
\def\rhat{\hat{r}}
\def\zhat{\hat{0}}
\def\kvec{{\mathbf k}}
\def\qvec{{\mathbf q}}
\def\qscaleV{\bm{\kappa}}
\def\qscaleS{{\kappa}}
\def\zedvec{{\mathbf z}}
\def\rvec{{\mathbf r}}
\def\Rvec{{\mathbf R}}
\def\udisvec{{\mathbf u}}
\def\udisscaF{U}
\def\udisvecF{{\mathbf U}}
\def\zvec{{\mathbf 0}}
\def\Dim{D}
\def\dim{d}
\def\efe{{\cal H}}
\def\pno{N}
\def\vol{V}
\def\ctp{\tau}
\def\hrs{\rm HRS}
\def\lrs{\rm LRS}
\def\cocon{g}
\def\locfrac{Q}
\def\dist{{\cal P}}
\def\scadist{\Pi}
\def\scat{\theta}
\def\sds{{\cal S}}
\def\hatpi{\hat{\Pi}}
\def\hatth{\hat{\scat}}
\def\tcomb{\mathbb T}
\def\barexi{\xi_{0}}
\def\ie{{\it i.e.\/}}
\def\eg{{\it e.g.\/}}
\def\viz{{\it viz.\/}}
\def\via{{\it via\/}}
\title{Scale-dependent elasticity as a probe of universal\\
heterogeneity in equilibrium amorphous solids}


\author{Boli Zhou}
\email[]{bzhou@utexas.edu}
\affiliation{Department of Physics, University of Texas at Austin,
Austin, TX 78712, USA}

\author{Rafael Hipolito}
\email[]{rafael.hipolito@stonybrook.edu}
\affiliation{Department of Physics, University of Texas at Austin,
Austin, TX 78712, USA}
\affiliation{Department of Physics and Astronomy, Stony Brook University,
Stony Brook, NY~11794, USA}

\author{Paul M.~Goldbart}
\email[]{paul.goldbart@stonybrook.edu}
\affiliation{Department of Physics and Astronomy, Stony Brook University,
Stony Brook, NY~11794, USA}


\date{January 2, 2023}

\begin{abstract}
The equilibrium amorphous solid state -- formed, \eg, by adequately randomly crosslinking the constituents of a macromolecular fluid -- is a heterogeneous state characterized by a universal distribution of particle localization lengths.
Near to the crosslink-density-controlled continuous amorphous-solidification transition, this distribution obeys a scaling form: it has a single peak at a lengthscale that diverges (along with the width of the distribution) as the transition is approached.
The modulus controlling macroscale elastic shear deformations of the amorphous solid does not depend on the distribution of localization lengths.
However, it is natural to anticipate that for deformations at progressively shorter lengthscales -- mesoscale deformations -- the effective modulus exhibits a scale-dependence, softening as the deformation lengthscale is reduced.
This is because an increasing fraction of the localized particles are, in effect, liquid-like at the deformation lengthscale, and therefore less effective at contributing to the elastic response.
In this paper, the relationship between the distribution of localization lengths and the scale-dependent elastic shear modulus is explored, and it is shown, within the setting of a replica mean-field theory, that the effective modulus does indeed exhibit scale-dependent softening.
Through this softening, mesoscale elasticity provides a probe of the heterogeneity of the state as characterized by the distribution of localization lengths.
In particular, the response to short-lengthscale elastic deformations is shown to shed light on the asymptotics of the universal localization-length distribution at short localization lengths.
\end{abstract}%



\maketitle

%
\section{\label{sec:intro}Introduction}
Fundamental to solids is the property of elasticity: responding to externally imposed, static, volume-preserving deformations of the shape of the solid by developing a static shear stress that resists the deformation (and {\it vice versa\/}).
The strength of this response can be characterized \via\ the static shear modulus: the larger this modulus, the larger is the shear stress required to produce a given deformation.
It is natural to extend this notion of elasticity away from the macroscale, to allow for position-dependent shear deformations and the associated position-dependent shear stresses that permeate the sample.
One can then ask: How does the elastic shear modulus change as the lengthscale of the deformation is tuned, from the (macroscopic) size of the sample, down in scale?
In a crystalline solid such as diamond or copper, for which crystallization from the molten state occurs \via\ a first-order phase transition, the relevant lengthscales describing the intrinsic fluctuations (and hence response) of the system remain microscopic throughout the solid state, all the way up in temperature to the melting point. Thus, the scale-dependent shear modulus is expected to vary only very weakly with the deformation scale, from the macroscale until atomic lengthscales begin to be approached, where a more substantial reduction in the modulus is expected.

Let us contrast the case of crystalline solids with vulcanized matter, such as that formed \via\ the crosslinking of macromolecules (see the foundational work by Deam and Edwards~\cite{D+Ephtl-1976} and, \eg, the discussions in Refs.~\cite{CGZaip-1996,PMGjop-2000}).
Two differences stand out.
First, the transition between the liquid state of lightly crosslinked systems and the solid state of well crosslinked systems is continuous, and is attended by a vanishing of the emergent shear modulus and a divergence in the spatial extent of the thermal motion exhibited by a typical localized particle, as the transition is approached from the solid side.
Second, rather than being crystalline, the state resulting from sufficient crosslinking is macroscopically homogeneous and isotropic and yet structurally inhomogeneous: 
the mean positions of the localized particles show no long-range periodicity and, furthermore, the localization lengths (\ie, RMS displacements) of the localized particles are spatially heterogeneous, varying randomly throughout the system.
The random structural heterogeneity can be captured most simply \via\ the statistical distribution of localization lengths presented by the system. This diagnostic turns out to be encoded in the order parameter that describes the amorphous solid state, and broadens and shifts to increasingly long lengthscales as the transition is approached. It is computable, at least at the level of mean-field theory.
At that level, and presumably beyond it, the distribution is found to be governed by a particular, parameter-free scaling function, which by a suitable rescaling accounts for all values of the distribution in the transition regime.
In this sense, the distribution of localization lengths is universal.

The amorphous solid state thus provides an unusual venue in which to explore scale-dependent elasticity. The continuous nature of the transition from the state, and the attendant divergence of the RMS displacements of the particles, suggest that, in contrast with crystalline solids, near to the transition the shear modulus should vary at the mesoscale, and not just at the microscale. Moreover, the heterogeneity of the state indicates that it is characterized, not by one or several lengths, but by a continuous family of lengths that range from the macroscale to the microscale.
The aim of this paper is to explore how these features of the equilibrium amorphous solid state -- localization length divergence and distribution -- are reflected in the scale-dependent elasticity and, conversely, to identify the extent to which scale-dependent elasticity presents an opportunity to probe the universal heterogeneity of the amorphous solid state.
\section{Ingredients\label{sec:ingredients}}
In this section, we collect the prior ideas and developments needed to address scale-dependent elasticity in the equilibrium amorphous solid state. We focus on the order parameter that detects and diagnoses this state as well as the effective Hamiltonian that governs the equilibrium order parameter and, hence, the state of the system, as a control parameter (such as the density of randomly introduced permanent constraints) is tuned. We review the structure of the amorphous solid state in preparation for discussing how the order parameter and free energy respond to elastic shear deformation at arbitrary lengthscales.
\subsection{Order parameter\label{sec:orderparameter}}
The order parameter field associated with the low-energy freedoms essential for addressing the transition is $\opf(\khat)$, whose argument is the $(1+n)$-fold replicated wave vector $\khat\equiv(\kvec^{0},\kvec^{1},\ldots,\kvec^{n})$; see Refs.~\cite{B+E-1980,GGprl-1987,CGZepl-1994,CGZaip-1996}. The $(1+n)$ vectors constituting $\khat$ are each $D$-dimensional, and are quantized appropriate to the periodic boundary conditions that are imposed on the replicas of the $D$-dimensional physical volume $\vol$ that contains the system. (The Fourier transform of $\opf(\khat)$ is real valued, as it corresponds to a joint probability distribution.)\thinspace\
Physically, $\opf$ is the analogue of the Edwards-Anderson spin-glass order parameter~\cite{E+Ajopl-1975} appropriate for identifying random localization. It serves as a detector and diagnoser of random heterogeneous localization, encoding the {\it fraction\/} $\locfrac$ of localized particles (which carries the percolative aspect of the transition) and the {\it distribution\/} $\dist(t)$ of inverse squared localization lengths $t$ of the localized particles (which captures the heterogeneity of the localization).
Respectively, its Fourier-space and real-space forms are:
\begin{subequations}
\begin{eqnarray}
\opf(\khat)&=&
\bigg[{1\over{N}}\sum_{j=1}^{N}
\prod_{\alpha=0}^{n}
\big\langle{\rm e}^{-i\kvec^{\alpha}\cdot\Rvec_{j}}\big\rangle\bigg],\\
\opf(\rhat)&=&
\bigg[{1\over{N}}\sum_{j=1}^{N}
\prod_{\alpha=0}^{n}
\big\langle\delta(\rvec^{\alpha}-\Rvec_{j})
\big\rangle\bigg],
\end{eqnarray}%
\end{subequations}%
where the $N$ particles are indexed by $j=1,\ldots,N$;
their instantaneous positions are given by $\Rvec_{j}$;
the angle brackets $\langle\cdots\rangle$ indicate thermal (\ie, annealed) averages; and the square brackets $[\cdots]$ indicate averages over crosslinking instances (\ie, quenched averages).
In a state in which all particles are delocalized, $\opf(\khat)=0$ (except at the trivial point $\khat=\zhat$); such states are invariant under {\it independent\/} translations of each of the $n$ replicas of the system, labelled $\alpha=0,\ldots,n$.
In a state in which at least some fraction of the particles are localized but there is no crystalline long-range order, $\opf(\khat)\ne 0$, but only for $\khat$ such that $\sum_{\alpha=0}^{n}\kvec^{\alpha}=\zvec$; such states are no longer invariant under independent translations of each of the replicas, but they retain the residual symmetry of invariance under {\it common\/} translations of the replicas. Macroscopically, the amorphous solid state retains the translation (and rotation) symmetry of the parent, liquid, state, but microscopically these symmetries spontaneously break at the transition.
\subsection{Effective Hamiltonian\label{sec:Hamilton}}%
We take for the effective Hamiltonian (or Landau-Wilson free energy; see Ref.~\cite{PCGZepl-1998}) $\efe[\opf]$, given by:
\begin{equation}
\efe=
\pno\sum_{\khat\in\hrs}\Big(
-\frac{a}{2}\ctp+
  \frac{\barexi^{2}}{2}
\khat^{2}\Big)
\vert\opf(\khat)\vert^{2}
-\pno\cocon
\!\!\!
\sum_{\khat_{1},\khat_{2},\khat_{3}\in\hrs}
\!\!\!
\delta_{\zhat,\khat_{1}+\khat_{2}+\khat_{3}}\,
\opf(\khat_{1})\,\opf(\khat_{2})\,\opf(\khat_{3}),
\label{eq:Landau-Wilson}
\end{equation}%
where $\ctp$ is the dimensionless control parameter for the transition and is determined, \eg, by the density of crosslinks that permanently constrain randomly selected segments of the system's macromolecules to remain adjacent to one another.
In addition, $\cocon$ measures the strength of the interactions between order-parameter fluctuations,
and $a$ and $\barexi$ are, respectively,
a number that characterizes the critical value of the crosslinking probability and a length that characterizes the bare size of the units that have been randomly connected (or, equally well, an effective lattice spacing). {\it Via\/} suitable rescalings
we may set $a/2$ and $g$ equal to unity, and we measure lengths in units of $\barexi$ and energies in units of the thermal energy scale $k_{\rm B}T$, where $T$ is the temperature.
We note that $\efe$ is invariant under independent translations of the replicas. The spontaneous breaking of this symmetry down to the symmetry of common translations marks the emergence of the amorphous solid state, as detected by the order parameter.

In expression~(\ref{eq:Landau-Wilson}) for $\efe$ and elsewhere, $\hrs$ stands for the {\it higher replica sector\/}, which means the collection of values of $\khat$ for which {\it at least two\/} of the replicated vectors in the set $\{\kvec^{0},\kvec^{1},\ldots,\kvec^{n}\}$ are nonzero. The significance of the $\hrs$ is that order-parameter components $\opf(\khat)$ attached to $\khat\in\hrs$ are critical degrees of freedom, which become soft near the solidification transition (or are continuously connected to such freedoms).
Complementary to the \hrs\ set of vectors is the \lrs\ set, where \lrs\ stands for {\it lower replica sector\/}. In contrast to $\khat\in\hrs$, interparticle repulsions strongly pin (to close to zero) the value of $\opf(\khat)$ attached to $\khat\in\lrs$. For this reason, it is appropriate to omit such freedoms from the description of the critical regime, either by constraining them to be zero or by integrating them out so as to (weakly) renormalize the parameters of the critical theory. (Trivially, $\opf(\khat)\vert_{\khat=\zhat}=1$ and does not fluctuate.)
\subsection{Equilibrium state\label{sec:equilib}}
The equilibrium-state value of $\opf$ (\viz, $\opfe$)
makes $\efe$ stationary (see Refs.~\cite{CGZepl-1994,PCGZepl-1998}):
\begin{equation}
0={\delta\efe\over{\delta\opf(-\khat)}}\Bigg\vert_{\opfe}=
2\Big(
-\ctp+{\khat^{2}\over{2}}
\Big)\,\opfe(\khat)-
3\!\!\!\sum_{\khat_{1},\khat_{2}\in\hrs}\!\!\!
\delta_{\khat,\khat_{1}+\khat_{2}}
\,\opfe(\khat_{1})\,\opfe(\khat_{2}).
\label{eq:statcon}
\end{equation}%
The liquid state (\ie, $\opfe=0$), stable at the classical level as a minimizer of $\efe$ provided $\ctp\le 0$, loses its stability for $\ctp>0$. Its place as the stable state is then taken by the following particular form:
\begin{equation}
\opf(\khat)=
\locfrac\,\delta_{\zvec,\tilde{\kvec}}
\int dt\,\dist(t)\,{\rm e}^{-\khat^{2}/2t},
\label{eq:opform}
\end{equation}%
where the integration over $t$ runs from $0$ to infinity (as it will all such integrals). Here, $\locfrac$ and $\dist$ are the number and function mentioned in Sec.~\ref{sec:intro}
as characteristics of the amorphous solid state,
\viz, the fraction of localized particles and the normalized distribution of their localization lengths (or, more precisely, of their inverse squared localization lengths). Furthermore, $\tilde{\kvec}\equiv\sum_{\alpha=0}^{n}\kvec^{\alpha}$ and
$\khat^{2}\equiv\sum_{\alpha=0}^{n}\kvec^{\alpha}\cdot\kvec^{\alpha}$. For an analysis of the stability of the random solid state, see Ref.~\cite{CGZepl-1999}.

It is convenient to exchange the unscaled variables $t$ and $\dist$ for the versions $\scat$ and $\scadist$ that are scaled in the following way by the control parameter $\ctp$:
\begin{subequations}
\begin{eqnarray}
t&=&(\ctp/2)\,\scat,\\
\dist(t)&=&(\ctp/2)^{-1}\,\scadist(\scat),
\end{eqnarray}%
\end{subequations}%
so that the scaled distribution $\scadist$ inherits normalization. Then, inserting the form~(\ref{eq:opform}) into the stationarity condition~(\ref{eq:statcon}),
evaluating the summations over $\khat_{1}$ and $\khat_{2}$ (restricted, importantly, to $\khat\in\hrs$),
and passing to the replica limit (\ie, taking $n\to 0$)
yields the consequences of stationarity for the physical parameters $\locfrac$ and $\scadist$ (or, equivalently, $\dist$):
\begin{subequations}
\begin{eqnarray}
0&=&
-2\,\ctp\,\locfrac+3\,\locfrac^{2},\\[4pt]
0&=&
{\scat^{2}\over{2}}{d\scadist\over{d\scat}}
-(1-\scat)\,\scadist(\scat)+(\scadist\circ\scadist)(\scat),
\label{eq:sceforscadist}%
\end{eqnarray}%
\label{eq:class-state}%
\end{subequations}%
where $\scadist\circ\scadist$ denotes the Laplace convolution of $\scadist$ with itself. [Note that Eq.~(\ref{eq:sceforscadist}) is consistent with $\scadist$'s being normalized.]\thinspace\ The transition to the amorphous solid state for $\ctp>0$ is marked by the emergence of a nonzero solution, $\locfrac=2\,\ctp/3$, to accompany the distribution that obeys Eq.~(\ref{eq:sceforscadist}).
\section{Elasticity\label{sec:disto}}
\subsection{Prior results
\label{sec:prior}}
The emergence of random localization of at least a fraction of its constituent particles, along with the development of solidness or rigidity characterized \via\ the advent of a zero-frequency elastic shear modulus, are two striking features of the equilibrium transition to the amorphous solid state exhibited by randomly constrained thermal systems such as vulcanized rubber. At least in spatial dimensions higher than two, these traits go hand in hand. The phenomenon of {\it macroscopic\/} elasticity has been examined from this order-parameter perspective, initially in Refs.~\cite{CGpre-1998,HECphd-1998,CGpre-2000} and then in Refs.~\cite{MGXZepl-2007,XMMphd-2008,MGXZpre-2009}, the latter focusing on the structure, implications, and interactions of Goldstone-type, low-wavelength, low-energy excitations of the amorphous solid order parameter. Whilst the aforementioned works focused on the transition regime, Ref.~\cite{UMGZepl-2006} examined macroscopic elasticity across the full range of crosslink densities. Here, we extend work on the elasticity of equilibrium amorphous solids beyond the macroscopic regime, bringing in mesoscopic elasticity and its implications by considering elastic shear deformations that vary in space on arbitrary lengthscales.
\subsection{Shear deformations
\label{sec:prior}}
We now recall the form of the order parameter when it is subject to a Goldstone-type deformation away from its equilibrium value;
see Refs.~\cite{MGXZepl-2007,XMMphd-2008,MGXZpre-2009}.
Bearing in mind the pattern of spontaneous symmetry breaking (\viz, from independent translations of the replicas down to common ones), one sees that the sector of low-energy deformations of the equilibrium order parameter is parametrized in terms of a set of $n$, position-dependent, $D$-vector-valued  displacement fields $\{\udisvec^{\alpha}(\zedvec)\}_{\alpha=1}^{n}$, reminiscent of a replicated elasticity theory.
Thus, the Goldstone-deformation
$\opfe+\delta\opf$
of the equilibrium order parameter
$\opfe$
can be expressed as follows:
\begin{subequations}
\begin{eqnarray}
\opfe(\khat)&=&
\locfrac\,\delta_{\zvec,\tilde{\kvec}}\,
\int dt\,\dist(t)\,{\rm e}^{-\khat^{2}/2t}
\\[2pt]
&=&
\locfrac
\int{d^{D}z\over{V}}\,
{\rm e}^{i\sum_{\alpha=0}^{n}\kvec^{\alpha}\cdot\zedvec}
\int dt\,\dist(t)\,{\rm e}^{-\khat^{2}/2t}
\\[2pt]
&&\longrightarrow
\locfrac\,
\int{d^{D}z\over{V}}\,
{\rm e}^{i\sum_{\alpha=0}^{n}\kvec^{\alpha}
\cdot\zedvec}\,
{\rm e}^{i\sum_{\alpha=1}^{n}\kvec^{\alpha}
\cdot\udisvec^{\alpha}(\zedvec)}
\int dt\,\dist(t)\,{\rm e}^{-\khat^{2}/2t}
\label{eq:deldefform}
\\[2pt]
&&\quad\approx
\opfe(\khat)+
\locfrac\,
\int{d^{D}z\over{V}}\,
{\rm e}^{i\sum_{\alpha=0}^{n}\kvec^{\alpha}\cdot\zedvec}\,
i\sum_{\alpha=1}^{n}\kvec^{\alpha}\cdot
\udisvec^{\alpha}(\zedvec)
\int dt\,\dist(t)\,{\rm e}^{-\khat^{2}/2t}
\\[2pt]
&&\quad\equiv\opfe(\khat)+\delta\opf(\khat),
\label{eq:deldef}
\end{eqnarray}%
\end{subequations}%
Consistent with the idea that common translations of the replicas remain symmetries of the amorphous solid state, there is one fewer displacement field than there are replicas; this is why there is no $\udisvec^{0}(\zedvec)$ present.
Consistent with the idea that a strong interparticle repulsion
heavily penalizes  fluctuations away from zero by the \lrs\ order-parameter fields, the displacement fields obey the incompressibility condition
${\rm det}
\big(
\delta_{dd^{\prime}}+
\partial u_{d}^{\alpha}(\zedvec)/
\partial z_{d^{\prime}}^{\phantom\alpha}
\big)=1$, which for small displacement gradients can be approximated by
$\partial u_{d}^{\alpha}(\zedvec)/
\partial z_{d}^{\phantom\alpha}=0$.
Said equivalently, we restrict our attention to shear deformations.
\subsection{Free-energy cost of shear deformations
\label{sec:costof}}
To determine the free-energy cost of elastic shear deformations in terms of the deformation fields $\{\udisvec^{\alpha}\}$, we observe [from Eqs.~(\ref{eq:deldefform}) and (\ref{eq:deldef})] that the order-parameter deformation $\delta\opf$ is, to leading order, linear in $\{\udisvec^{\alpha}\}$. With this in mind, we expand $\efe[\opf]$, Eq.~(\ref{eq:Landau-Wilson}), around the equilibrium state $\opfe$, keeping terms up to second order in the deformation $\delta\opf$ and hence in $\{\udisvec^{\alpha}\}$. However, we forgo the customary additional \lq\lq gradient\rq\rq\ expansion in powers of the wave vector of the deformation, instead retaining the full wave-vector dependence, so that we may identify -- as fully as is possible within the present (Landau, or classical, or mean-field) scheme -- the dependence of the elastic shear modulus on the lengthscale of the elastic deformation.

Implementing this scheme (and omitting the term linear in $\delta\opf$, as it vanishes by virtue of the stationarity condition on $\opf$), we have:
\begin{subequations}
\begin{eqnarray}
&&
\!\!\!\!
\delta\efe\big[\,\opfe,\udisvec\,\big]
\equiv
\efe\big[\,\opfe+\delta\opf(\udisvec)\big]-
\efe\big[\,\opfe\,\big]
\approx
{1\over{2}}\!
\sum_{\khat_{1},\khat_{2}\in\hrs}
{\delta^{2}\efe\over{\delta\opf(\khat_{2})\delta\opf(\khat_{1})}}
\Bigg\vert_{\opfe}
\delta\opf(\khat_{1})\,\delta\opf(\khat_{2})
\\
&&
=\pno\sum_{\khat\in\hrs}
\Big(
-\ctp+
\frac{\khat^{2}}{2}\Big)
\vert\delta\opf(\khat)\vert^{2}
-3\pno
\!\!\!
\sum_{\khat_{1},\khat_{2},\khat_{3}\in\hrs}
\!\!\!
\delta_{\zhat,\khat_{1}+\khat_{2}+\khat_{3}}\,
\opfe(\khat_{3})\,\delta\opf(\khat_{1})\,\delta\opf(\khat_{2})
\nonumber
\\
&&
\approx
\pno\sum_{\khat\in\hrs}
\Big(-\ctp+\frac{\khat^{2}}{2}\Big)
\nonumber
\\
&&\qquad\qquad
\times\,\locfrac
\int dt_{1}\,\dist(t_{1})\,
{\rm e}^{-\khat^{2}/2t_{1}}
\int{d^{D}z_{1}\over{V}}\,
{\rm e}^{-i\sum_{\beta_{1}=0}^{n}
\kvec^{\beta_{1}}\cdot\,\zedvec_{1}}\,
(-i)\sum_{\alpha_{1}=1}^{n}\kvec^{\alpha_{1}}
\cdot\udisvec^{\alpha_{1}}(\zedvec_{1})
\nonumber
\\
&&\qquad\qquad
\times\,\locfrac
\int dt_{2}\,\dist(t_{2})\,
{\rm e}^{-\khat^{2}/2t_{2}}
\int{d^{D}z_{2}\over{V}}\,
{\rm e}^{+i\sum_{\beta_{2}=0}^{n}
\kvec^{\beta_{2}}\cdot\,\zedvec_{2}}\,
(+i)\sum_{\alpha_{2}=1}^{n}\kvec^{\alpha_{2}}
\cdot\udisvec^{\alpha_{2}}(\zedvec_{2})
\nonumber
\\
&&\quad-\,3\pno
\!\!\!
\sum_{\khat_{1},\khat_{2},\khat_{3}\in\hrs}
\!\!\!
\delta_{\zhat,\khat_{1}+\khat_{2}+\khat_{3}}\,
\locfrac
\int dt_{3}\,\dist(t_{3})\,
{\rm e}^{-\khat_{3}^{2}/2t_{3}}
\int{d^{D}z_{3}\over{V}}\,
{\rm e}^{-i\sum_{\beta_{3}=0}^{n}
\kvec_{3}^{\beta_{3}}\cdot\,
\zedvec_{3}^{\phantom{\beta_{3}}}}
\nonumber
\\
&&\qquad\qquad
\times\,\locfrac
\int dt_{1}\,\dist(t_{1})\,
{\rm e}^{-\khat_{1}^{2}/2t_{1}}
\int{d^{D}z_{1}\over{V}}\,
{\rm e}^{-i\sum_{\beta_{1}=0}^{n}
\kvec_{1}^{\beta_{1}}\cdot\,\zedvec_{1}^{\phantom{\beta_{1}}}}\!
(-i)\sum_{\alpha_{1}=1}^{n}\kvec_{1}^{\alpha_{1}}
\cdot\udisvec^{\alpha_{1}}(\zedvec_{1})
\nonumber
\\
&&\qquad\qquad
\times\,\locfrac
\int dt_{2}\,\dist(t_{2})\,
{\rm e}^{-\khat_{2}^{2}/2t_{2}}
\int{d^{D}z_{2}\over{V}}\,
{\rm e}^{+i\sum_{\beta_{2}=0}^{n}
\kvec_{2}^{\beta_{2}}\cdot\,\zedvec_{2}^{\phantom{\beta_{2}}}}\!
(+i)\sum_{\alpha_{2}=1}^{n}\kvec_{2}^{\alpha_{2}}
\cdot\udisvec^{\alpha_{2}}(\zedvec_{2}).
\label{unevalform}%
\end{eqnarray}%
\end{subequations}%
This formula gives the free-energy cost of elastic shear deformations: to second order in the dependence on the displacement fields and to all orders in the dependence on the wavelength content of the displacement fields. What remains is to evaluate the formula as fully as is possible, including taking the replica limit and recognizing the translational invariance of the resulting (pure, effective, displacement-field-dependent) elastic free-energy.
\subsection{Scale-dependent shear modulus
\label{sec:sceledep}}
To determine scale-dependent elastic modulus associated with elastic shear deformations, we evaluate $\delta\efe\big[\,\opfe,\udisvecF\,\big]$ using Eq.~(\ref{unevalform}), by inserting the properties of the equilibrium state given in Eqs.~(\ref{eq:class-state}), including the scaling behavior of the state with the control parameter $\ctp$. Thus, we arrive at:
\begin{subequations}%
\begin{equation}%
\delta\efe\big[\,\opfe,\udisvecF\,\big]
=
-{\pno\over{2\vol}}
\sum_{\underset{(\alpha_{1}\,\ne\,\alpha_{2})}
        {\alpha_{1},\,\alpha_{2}\,=\,1}}^{n}
{1\over{\vol}}\sum_{\qvec}\,
\sds(\qvec)\,\vert\qvec\vert^{2}\,
\udisvecF^{\alpha_{1}}(\qvec)^{\ast}\cdot
\udisvecF^{\alpha_{2}}(\qvec),
\end{equation}%
where the wave-vector-dependent shear modulus $\sds(\qvec)$,
its long-distance limit $\sds(\zvec)$, and the dimensionless
scaling function $\Sigma(\kappa)$ (which is completely
determined by $\scadist$) are given by:
\begin{eqnarray}%
\sds(\qvec)
&\equiv&
\sds(\bm{0})\,
\Sigma(\vert\qvec\vert^{2}/\ctp),\quad{\rm in\,\, which}
\\[2pt]
\sds\big(\bm{0}\big)
&\equiv&
k_{\rm B}T\,
(4\,\ctp^{3}/27),\quad{\rm and}
\\[5pt]
\Sigma(\kappa)
&\equiv&
{3\over{4\kappa}}
\Big(
4\big\{
\tcomb_{12}\,
{\rm e}^{-\kappa/\tcomb_{12}}
\big\}
+2\big\{
\tcomb_{12}^{2}\,
{\rm e}^{-\kappa/\tcomb_{12}}
\big\}
\nonumber\\
&&
\quad\,
-8\big\{
\big(\tcomb_{12}^{-1}+(\scat_{3}/\scat_{1}\scat_{2})\big)^{-1}\,
{\rm e}^{-\kappa/\tcomb_{12}}
\big\}
\Big)
+{3\over{2}}
\big\{
\tcomb_{12}\,
{\rm e}^{-\kappa/\tcomb_{12}}
\big\}.
\label{scaled-dist}
\end{eqnarray}%
\label{eq:setof}%
\end{subequations}%
Here, for the sake of compactness, we have introduced the symmetric combination of variables $\tcomb_{12}$, defined \via\
$\tcomb_{12}^{-1}\equiv\big(\scat_{1}^{-1}+\scat_{2}^{-1}\big)$.
Similarly, we have introduced the braces notation $\{\cdots\}$ to indicate averaging over the (random, inverse-squared, scaled) localization lengths $\scat_{1}$, $\scat_{2}$ etc.:
$$\{\cdots\}=
\int d\scat_{1}\,\scadist(\scat_{1})
\int d\scat_{2}\,\scadist(\scat_{2})
\int d\scat_{3}\,\scadist(\scat_{3})\cdots .$$
In addition, we have exchanged the displacement fields
$\{\udisvec^{\alpha}\}_{\alpha=1}^{n}$
for their Fourier transforms
$\{\udisvecF^{\alpha}\}_{\alpha=1}^{n}$
defined \via\ the pair:
\begin{subequations}
\begin{eqnarray}
\udisvecF(\qvec)
&=&
\int{d^{D}z}\,
{\rm e}^{i\qvec\cdot\zedvec}\,
\udisvec(\zedvec),
\\
\udisvec(\zedvec)
&=&
{1\over{\vol}}\sum_{\qvec}
{\rm e}^{-i\qvec\cdot\zedvec}\,
\udisvecF(\qvec).
\end{eqnarray}%
\end{subequations}%
That the dimensionless scaling function $\Sigma(\kappa)$ is determined by $\scadist$ is the key result of this paper, as it reveals how the scale-dependence of the elastic shear modulus can serve as a probe of the distribution of localization lengths.
\section{Characteristics of the scale-dependent shear modulus
\label{sec:chara}}
\subsection{Numerical results at all scales
\label{sec:allchara}}
\begin{figure}
\centering
\includegraphics[width=0.8\textwidth]{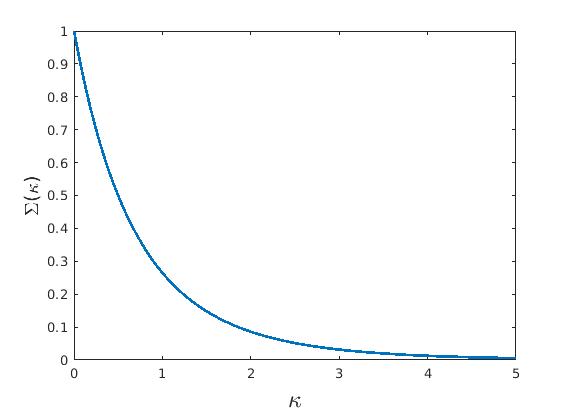}
\caption{Dimensionless scale-dependent shear modulus
$\Sigma(\kappa)$, defined in Eq.~(\ref{scaled-dist}).
\label{fig:sigma}}
\end{figure}

Figure~\ref{fig:sigma} shows the dimensionless scaling function $\Sigma$ as a function of $\kappa$, computed numerically using Eq.~(\ref{scaled-dist}) in terms of the numerically computed dimensionless distribution $\scadist$ (known from Ref.~\cite{CGZepl-1994}).
For the expected reasons that are discussed in Sec.~\ref{sec:limitchara} below, $\Sigma_{0}=1$, independent of the form $\dist$.
As the figure shows, the dimensionless scaling function $\Sigma(\kappa)$ decreases monotonically as $\kappa$ is increased (\ie, as the lengthscale of the elastic shear deformation is decreased), and it tends to zero for large $\kappa$ (\ie, for deformations of small wavelength). The change is most rapid when $\kappa={\cal O}(1)$, i.e., when the deformation lengthscale $2\pi/\vert\qvec\vert$ is comparable to the typical localization length, i.e., $\xi_{0}/\sqrt{\ctp}$. This behavior reflects the physical idea that the shorter the lengthscale of the deformation, the smaller is the fraction of particles that are sufficiently well-localized to contribute potently to the elastic response.
Particles that are localized on lengthscales rather longer than the deformation lengthscale contribute less efficaciously towards the elasticity of the medium, because from the viewpoint of the deformation lengthscale they appear liquid-like. This is why $\Sigma$ changes most rapidly at a deformation lengthscale that is comparable to the most probable localization length.
\subsection{Large-distance behavior, first deviations,
        and their implications
\label{sec:limitchara}}
Returning to analytics, we first confirm that the form of the behavior of $\sds(\qvec)$ at asymptotically
long lengthscales is indeed that reported in Eqs.~(\ref{eq:setof}), \ie, that
$\lim_{\kappa\to 0}\Sigma(\kappa)=1$.
To check that this is the case, we examine Eq.~(\ref{scaled-dist}), which for $\vert\kappa\vert\ll 1$ gives:
\begin{subequations}
\begin{eqnarray}
\Sigma(\kappa)
&=&
\kappa^{-1}\,\Sigma_{-1}+
\kappa^{ 0}\,\Sigma_{0}+
\kappa^{ 1}\,\Sigma_{1}+
{\cal O}\big(\kappa^{2}\big),\quad{\rm where}
\label{eq:sigtot}
\\[6pt]
\Sigma_{-1}
&\equiv&
{3\over{4}}
\Big(
4\big\{
\tcomb_{12}
\big\}
+2\big\{
\tcomb_{12}^{2}
\big\}
-8\big\{
\big(\tcomb_{12}^{-1}+(\scat_{3}/\scat_{1}\scat_{2})\big)^{-1}
\big\}
\Big),
\label{eq:sigminus}
\\[6pt]
\Sigma_{0}
&\equiv&
-\,3+6\,\big\{
\tcomb_{12}^{-1}\,
\big(\tcomb_{12}^{-1}+(\scat_{3}/\scat_{1}\scat_{2})\big)^{-1}
\big\},
\label{eq:sigzero}
\\[6pt]
\Sigma_{1}
&\equiv&
-{3\over{4}}\,
\big(1+
4\big\{(\scat_{1}+\scat_{2}+\scat_{3})^{-1}\big\}
\big),
\label{eq:sigpone}
\end{eqnarray}%
\end{subequations}%
$\Sigma_{-1}$, which controls the ${\cal O}(\kappa^{-1})$ term, vanishes. This follows from the particular form of the equilibrium value of $\dist$, which we invoke \via\ the equation obeyed by its Laplace transform:
\begin{equation}
\hatth\,{d^{2}\hatpi\over{d\hatth^{2}}}=2\,(1-\hatpi)\,\hatpi.
\label{eq:LTdistrib}
\end{equation}%
Examining in turn the three terms in Eq.~(\ref{eq:sigminus}) for $\Sigma_{-1}$, we have:
\begin{subequations}%
\begin{eqnarray}%
&&
\big\{\tcomb_{12}\big\}=
\int d\scat_{1}\,\scat_{1}\,\pi(\scat_{1})\,
     d\scat_{2}\,\scat_{2}\,\pi(\scat_{2})
\int d\hatth\,{\rm e}^{-\hatth(\scat_{1}+\scat_{2})}=
\int d\hatth\,\hatpi^{\prime}(\hatth)^{2},\\
&&
\big\{\tcomb_{12}^{2}\big\}=
\int d\scat_{1}\,\scat_{1}^{2}\,\pi(\scat_{1})\,
     d\scat_{2}\,\scat_{2}^{2}\,\pi(\scat_{2})
\int d\hatth\,\hatth\,{\rm e}^{-\hatth(\scat_{1}+\scat_{2})}=
\int d\hatth\,\hatth\,\hatpi^{\prime\prime}(\hatth)^{2},\\
&&
\big\{
\big(\tcomb_{12}^{-1}+(\scat_{3}/\scat_{1}\scat_{1})\big)^{-1}
\big\}=
\int d\scat_{1}\,\scat_{1}\,\pi(\scat_{1})\,
     d\scat_{2}\,\scat_{2}\,\pi(\scat_{2})\,
     d\scat_{3}\,           \pi(\scat_{3})
\int d\hatth\,{\rm e}^{-\hatth(\scat_{1}+\scat_{2}+\scat_{3})}
\\
\nonumber
&&\qquad\qquad\qquad\qquad\qquad=
\int d\hatth\,\hatpi(\hatth)\,
            \hatpi^{\prime}(\hatth)^{2},
\end{eqnarray}%
\end{subequations}%
where all integrations run from $0$ to $\infty$.
By substituting these integral representations for the three terms in Eq.~(\ref{eq:sigminus}), we obtain:
\begin{equation}
\Sigma_{-1}=
{3\over{2}}\int d\hatth\,
\big(\,2(\hatpi^{\prime})^{2}+
 \hatth(\hatpi^{\prime\prime})^{2}-
4\hatpi\,(\hatpi^{\prime})^{2}\,\big).
\end{equation}%
Eliminating the second-derivative term, using  equation~(\ref{eq:LTdistrib}) obeyed by $\hatpi$, 
transforms the integrand into a total derivative, 
which may be integrated to give:
$\Sigma_{-1}=
3(1-\hatpi)\,\hatpi\,
\hatpi^{\prime}\big\vert_{0}^{\infty}\,$. 
This vanishes as a consequence of the boundary conditions
$\hatpi(\hatth)\vert_{\hatth=0}=1$ and $\hatpi(\hatth)\vert_{\hatth=\infty}=0$; 
hence, as required, $\Sigma_{-1}=0$.

Turning now to $\Sigma_{0}$, the following elementary manipulation reveals that $\Sigma_{0}=1$. Note that
$\big\{
\tcomb_{12}^{-1}
\big(\tcomb_{12}^{-1}+(\scat_{3}/\scat_{1}\scat_{1})\big)^{-1}
\big\}=
(1/3)
\left\{
\left({(\scat_{1}+\scat_{2}+\scat_{2}+\scat_{3}+\scat_{3}+\scat_{1})/
       (\scat_{1}+\scat_{2}+\scat_{3}})\right)
\right\}
=2/3$;
hence $\Sigma_{0}=1$, independent of the form $\dist$, as one expects (and as was already recognized in Refs.~\cite{CGpre-1998,CGpre-2000}), given that at asymptotically long lengthscales all localized particles {\it look\/} perfectly localized, so all that should determine the elastic modulus in this limit is the fraction of particles that are localized at all.

As for $\Sigma_{1}$, by invoking the standard representation
$\scat^{-1}=\int_{0}^{\infty}d\hatth\,{\rm e}^{-\hatth\scat}$,
it is straightforward to see that: 
\begin{equation}
\Sigma_{1}=
-{3\over{4}}
\Big(1+
4\int d\hatth\,\hatpi(\hatth)^{3}
\Big).
\end{equation}%
As expected, regarding the progression from the largest to shorter lengthscales, this is the first place where sensitivity to the distribution of localization lengths enters. It expresses the initial weakening of the elasticity as the probe scale approaches the typical localization-length scale from above, and occurs because particles characterized by localization lengths longer than the probe scale are less effective at producing an elastic response.
What controls the strength of this effect is
$\big\{(\scat_{1}+\scat_{2}+\scat_{3})^{-1}\big\}$,
which amplifies the influence of the more weakly localized fraction. Equivalently, since $\hatpi$ decays monotonically from $1$, with increasing $\hatth$ (owing to the normalization and non-negativity of the probability distribution $\scadist$), the third power of $\hatpi$ reinforces the decay, thus emphasizing the role of weakly localized particles.
Putting the pieces together, for small
$\xi_{0}^{2}\,\vert\qvec\vert^{2}/\ctp$ we have:
\begin{equation}
\sds(\qvec)
\approx
k_{\rm B}T\,
{4\,\ctp^{3}\over{27}}\,
\bigg(1-
{3\over{4}}\,
{\xi_{0}^{2}\,\vert\qvec\vert^{2}\over{\ctp}}
\Big(1+4\int d\hatth\,\hatpi(\hatth)^{3}\Big)+
{\cal O}\big(\xi_{0}^{4}\,\vert\qvec\vert^{4}/\ctp^{2}\,\big)
\bigg).
\end{equation}%
It is straightforward to obtain the
${\cal O}\big(\,\xi_{0}^{4}\,\vert\qvec\vert^{4}/\ctp^{2}\big)$
correction to $\sds(\qvec)$ as well as higher-order corrections, and to ascertain their sensitivity to $\scadist$.
\subsection{Small-distance behavior and implications
\label{sec:shortchara}}
Lastly, we take this perspective to the other extreme of lengthscales, enquiring about the asymptotic behavior of $\sds(\qvec)$ at lengthscales much shorter than the characteristic localization length, \ie, for $\vert\qvec\vert^{2}\gg\ctp$.
Thus, we consider equation~(\ref{scaled-dist}) for $\Sigma(\kappa)$ for $\kappa\gg 1$, and apply Laplace's method with movable
maxima~(see, e.g., Ref.~\cite{B+Oamm-1999}, Sec.~6.4)
to the separate determination of the asymptotic behavior of each of the four terms that constitute $\Sigma(\kappa)$.
One finds that the dominant contribution comes from the last of the four, so we give details for that term only. Thus, we consider:
\begin{equation}
{\mathbb A}(\kappa)
\equiv
\big\{\tcomb_{12}\,{\rm e}^{-\kappa/\tcomb_{12}}\big\},
\end{equation}%
and observe that
\begin{equation}
-{d{\mathbb A}\over{d\kappa}}=
\big\{
{\rm e}^{-\kappa/\tcomb_{12}}
\big\}=
\big\{{\rm e}^{-{\kappa/\scat}}\big\}^{2}.
\end{equation}%
Focusing, then, on $\{{\rm e}^{-{\kappa/\scat}}\}$, we note that at large $\kappa$ the integral over $\scat$ is dominated by the large-$\scat$ regime. As a result, we may replace $\Pi(\scat)$ by its large-$\scat$ form, which we know from
Refs.~\cite{HECphd-1998,CGZepl-1994} to be given by:
\begin{subequations}
\begin{eqnarray}
\Pi(\scat)&\approx&3\,b\,\scat\,{\rm e}^{-b\scat}\quad({\rm for}\,\,\scat\gg 1),\\
b&\approx&1.678.
\end{eqnarray}%
\end{subequations}%
Thus, we arrive at: 
\begin{equation}
\big\{{\rm e}^{-{\kappa/\scat}}\big\}=
\int d\scat\,\Pi(\scat)\,
{\rm e}^{-{\kappa/\scat}}
\approx
3b\int d\scat\,
\scat\,{\rm e}^{-b\scat}\,
{\rm e}^{-{\kappa/\scat}}.
\end{equation}%
At large $\kappa$, the maximum of the integrand occurs at $\bar{\scat}$ obeying: 
\begin{equation}
{d\over{d\scat}}
\left(
-b\scat-{\kappa\over{\scat}}
\right)\bigg\vert_{\scat=\bar{\scat}}=0,
\end{equation}%
\ie, $\bar{\scat}=\sqrt{\kappa/b\,}$, which moves with $\kappa$.
Next, we exchange the integration variable $\scat$ for $y$, {\it via\/} $\scat=\bar{\scat}(1+y)$, where $y=0$ is now the maximizer of the exponent.
Expanding the exponent about $y=0$, we obtain for it, to sufficient accuracy:
$-2\sqrt{b\kappa}-\sqrt{b\kappa\,}y^{2}$.
We then expand the non-exponential factor around $y=0$ and, as always for Laplace methods with an interior maximum, we extend the integration range for $y$ to the complete real line and then perform the resulting Gaussian integration, in this case obtaining:
\begin{equation}
-{d\over{d\kappa}}{\mathbb A}(\kappa)\approx
9\pi\, b^{-1/2}\kappa^{3/2}\,{\rm e\/}^{-4\sqrt{b\kappa}}.
\end{equation}%
Integrating with respect to $\kappa$, recognizing that
${\mathbb A}(\infty)=0$,
we find: 
\def\kapint{\bar{\kappa}}
\begin{equation}
{\mathbb A}(\kappa)=
-\big({\mathbb A}(\infty)-{\mathbb A}(\kappa)\big)=
-\int_{\kappa}^{\infty}d\kapint\,
{d{\mathbb A}\over{d\kapint}}
\,\approx\,
{9\pi\over{b^{1/2}}}
\int_{\kappa}^{\infty}d\kapint\,
\kapint^{3/2}\,
{\rm e\/}^{-4\sqrt{b\kapint}}.
\end{equation}%
Because we are concerned with large $\kappa$, we may apply Laplace's method with a maximum at the boundary, in this case 
the lower
boundary (see, e.g., Ref.~\cite{B+Oamm-1999}, Sec.~6.4),
thus arriving at the sought asymptotic behavior of ${\mathbb A}$,
\begin{equation}
{\mathbb A}(\kappa)
\approx
{9\pi\over{2b}}\,\kappa^{2}\,
{\rm e\/}^{-4\sqrt{b\kappa}}\quad({\rm for}\,\,\kappa\gg 1),
\end{equation}%
and hence of $\Sigma(\kappa)$,
\begin{equation}
\Sigma(\kappa)
\approx
{27\over{4}}{\pi\over{b}}\,\kappa^{2}\,
{\rm e\/}^{-4\sqrt{b\kappa}}\quad({\rm for}\,\,\kappa\gg 1).
\end{equation}%
Thus, from Eqs.~(\ref{eq:setof}), we arrive at the leading large-$\vert\qvec\vert$ behavior of the scale-dependent elastic shear modulus:
\begin{equation}
\sds(\qvec)\approx
k_{\rm B}T\,
\big(\pi/b\big)\,\ctp^{3}\,
\big(\,\xi_{0}^{2}\,\vert\qvec\vert^{2}/\ctp\big)^{2}\,
\exp\!\big({-4\sqrt{b\,\xi_{0}^{2}\,\vert\qvec\vert^{2}/\ctp}}\big)
\quad({\rm for}\,\,\xi_{0}^{2}\,\vert\qvec\vert^{2}\gg\ctp).
\end{equation}%
Note the exponential dependence on $\vert\qvec\vert$ (and not $\vert\qvec\vert^{2}$).
This reflects two significant qualitative features of the amorphous solid state:
(i)~that the localization lengths are continuously distributed; and
(ii)~that the support of the distribution is not bounded below. In other words, the distribution has weight that reaches down to arbitrarily small localization lengths, a feature that puts the amorphous solid state is a class of its own, qualitatively distinguished from the class of pure equilibrium solids, such as diamond or copper.
\section{Concluding remarks
\label{sec:conclude}}
In this paper, we have applied replica mean-field theory to the topic of the elasticity of equilibrium amorphous solids. We have focused on the variation of the elastic shear modulus with the lengthscale of the corresponding shear deformation. We have demonstrated that, in view of the continuous nature of the transition to the amorphous solid state and the heterogeneity of the structure of the emergent state, in the transition regime the shear modulus should display striking scale dependence, not just at the microscopic scale of atoms and molecules, but even at the emergent, collective, mesoscopic scale of position fluctuations of localized entities. We have, furthermore, discussed the origins of this scale dependence in terms of the diminution of the contribution to the elasticity of particles that are localized on scales longer than a given elastic deformation lengthscale. This interconnection between the scale-dependent elasticity and the structural heterogeneity (as characterized by the distribution of localization lengths) opens up the possibility of using scale-dependent elasticity as a probe of the qualitative and even quantitative nature of the distribution of localization lengths. It would be interesting to understand the extent to which the ideas presented here survive the incorporation of order-parameter fluctuations, or can at least be couched in terms of the kinds of asymptotic, beyond-mean-field-theory scaling variables and functions that a renormalization-group analysis could, in principle, determine.
\begin{acknowledgments}
The work of PMG was performed in part at the Aspen Center for Physics, which is supported by National Science Foundation grant PHY-1607611.
\end{acknowledgments}

{\raggedright 

} 
\bibliography{basename of .bib file}

\bibliography{sources}

\end{document}